\documentclass[aps,superscriptaddress,twocolumn,showpacs,preprintnumbers,amsmath,amssymb]{revtex4}
\usepackage[cp1251]{inputenc}
\usepackage{amssymb,amsmath}
\usepackage[mathscr]{eucal}
\usepackage[pdftex]{graphicx}
\usepackage{longtable}
\usepackage[unicode, colorlinks=false,linkcolor=blue, pdfborder={0 0 0}]{hyperref} 
\begin{document}
\title{Symmetry Properties of Electromagnetic Field in the Matter}
\author{D.Yearchuck}
\affiliation{Minsk State Higher Aviation College, Uborevich Str.77, Minsk, 220096, yearchuck@gmail.com} 
\author{A.Alexandrov}
\affiliation{M.V.Lomonosov Moscow State University, Moscow, 119899, alex@ph-elec.phys.msu.su}
             
\begin{abstract} The sets ${\Phi({F}^{\mu \nu})}, {\Phi(\tilde {F}^{\mu \nu})}$ of linear functionals on the space $\left\langle F,+,\cdot \right\rangle$ represent themself linear space $\left\langle \Phi,+,\cdot \right\rangle$ over the field of \textit{scalars} $P$, which is dual to space $\left\langle F,+,\cdot \right\rangle$, but it is substantial, that given linear space is not self-dual. It has been found, that the partition of
linear space $\left\langle F,+,\cdot \right\rangle$ over the field of genuine scalars and pseudoscalars, the vectors in which are sets of contravariant and covariant electromagnetic field tensors and pseudotensors $\left\{{F}^{\mu\nu}\right\}$, $\left\{\tilde{F}^{\mu\nu}\right\}$, $\left\{{F}_{\mu\nu}\right\}$, $\left\{\tilde{F}_{\mu\nu}\right\}$, on 4 subspaces takes place. It corresponds to appearance of 4 kinds of electromagnetic field potential 4-vectors $A_{\mu}$, which are transformed according to the representations of general Lorentz group with various symmetry relatively improper rotations. It has been found, that conserving quantity, corresponding to complex fields is complex charge. It is argued, that electromagnetic field in the matter is complex field and 
 that two-parametric group $\Gamma(\alpha,\beta) = U_{1}(\alpha) \otimes \mathfrak R(\beta)$, where $\mathfrak R(\beta)$ is abelian multiplicative group of real numbers (excluding zero), determines the gauge symmetry of electromagnetic field. It is also argued, that free electromagnetic field is characterized by pure imagine charge. 
\end{abstract} 

\maketitle 
\section{Introduction and Background}
The operator equation, describing the optical transition dynamics, has recently been obtained by using of transition operator method \cite{Yearchuck_Doklady}. It has been shown, that given equation is operator equivalent to Landau-Lifshitz (L-L) equation \cite{Landau} in its 
difference-differential form, which takes usual differential form in continuum limit. In view of isomorphism of algebras of 
transition operators $\hat {\vec {\sigma }}_k $ and components of the spin S = 1/2, the symmetry of Bloch vector $\vec {P}$ in optical Bloch equations relatively improper rotations and the physical meaning of all its components have been established. Let remember, that optical Bloch equations are the essence of gyroscopic model for spectroscopic transitions proposed for the first time formally \cite{Abragam} by
F.Bloch. So, it was concluded in \cite{Yearchuck_Doklady}, that Bloch vector is electric dipole moment, in particular, electric spin moment. It has to be identical in its symmetry properties relatively improper rotations to vector of spin moment, that is, it is axial vector. 
The nature of the second vector, that is, $\vec {E}$, entering
optical Bloch equations, has also been clarified. It is electric field vector. The mathematical structure of L-L equation allowed to establish unambiguously the symmetry of electric field vector relatively improper rotations to be also axial vector. Given conclusion and the model used at all were experimentally confirmed in \cite{Yearchuck_Yerchak}, \cite{Yearchuck_PL} by means of observation of ferroelectric \cite{Yearchuck_Yerchak} and antiferroelectric \cite{Yearchuck_PL} spin wave resonances, which were predicted on the base of the model \cite{Yearchuck_Doklady} for the chain of electrical spin moments, interacting between themselves by a mechanism like to Heisenberg exchange. Especially interesting, that in \cite{Yearchuck_PL} was experimentally proved, that really pure imagine electrical spin moment, in full correspondence with Dirac prediction \cite{Dirac1928}, is responsible for the phenomenon observed. Electric dipole moment seems to be including in general case along with a spin part an orbital part, which is, evidently, a real-valued quantity. 

Although the conclusion on unusual symmetry properties of electrical characteristics of electromagnetic field (EM-field), based on the comparison of the theory of spectroscopic transitions, proposed in \cite{Yearchuck_Doklady}, with the experimental data above indicated, was done, the cause of given phenomenon cannot be explained in the frames of the only spectroscopic transitions' theory. Really, it has to be explained in fact the phenomenon of dual symmetry of EM-field vectors, that is, the appearance of axial properties of electrical vectors by spectroscopic transitions, while, for instance, electric charge transport properties can be well described in the suggestion, that electric field vector is polar vector. Consequently, additional general considerations of the symmetry properties of EM-field characteristics and field theory consideration at all are needed. The known field theory consideration allows to explain the only one possibility: 4-vector of EM-potential has to be polar and t-even 4-vector \cite{Akhiezer}. It means, that electric field vector has to be, according to given possibility, the only polar vector and magnetic field vector has to be the only axial vector. So, the first task of presented paper is to study the symmetry properties of EM-field characteristics relatively improper rotations in more details.

The second task follows from the comparison of experimental results, reported in \cite{Ertchak_J_Physics_Condensed_Matter}, \cite{Ertchak_Carbyne_and_Carbynoid_Structures} and \cite{Yearchuck_Yerchak}. Carbynoid samples, studied in given works, were active both in magnetic resonance and optical infrarot (IR) and Raman scattering (RS) spectroscopy. Carbynoid samples, representing themselves the systems of carbon chains, can be considered to be the most simple modelling systems for verification of theoretically predicted effects for quasionedimensional structures. Among the samples studied were the samples (designated samples A in \cite{Yearchuck_Yerchak}), in which the ferromagnetic and ferroelectric spin wave resonances (FMSWR and FESWR) were observed on the same chain structures. Both the resonances can be described by in fact the same equation, obtained in \cite{Yearchuck_Doklady} and modified by taking into consideration 
the relaxation processes \cite{Yearchuck_Yerchak}. So, for description of optical transition dynamics the equation 
is: 
\begin{equation}
\label{eq1}
\begin{split}
\raisetag{40pt}
\frac{\partial \vec {S}(z)}{\partial t} = \left[ {\vec {S}(z)\times \gamma_{E} 
\vec {E}} \right] - \frac{4a^2J_{E}}{\hbar ^2}\left[ {\vec {S}(z)\times \nabla 
^2\vec {S}(z)} \right] \\+ \frac{\vec {S}(z)-\vec {S}_0 (z)}{\tau },
\end{split}
\end{equation}
where $\vec {S}(z)$ is electric analogue of spin magnetic moment, $\gamma_{E}$, $J_{E}$ are optical analogues of gyromagnetic ratio and exchange 
interaction constant respectively, $\hbar$ is Planck's constant, $a$ is lattice 
spacing, $\vec {E}$ is electric field, 
 $\vec {S}_0 (z)$ is equilibrium value of electrical spin moment 
vector function, $\tau$ is relaxation time. 
Vector-functions $\vec {S}(z)$ $\vec {S}_0 (z)$ acquire in the case of FMSWR the meaning of magnetic spin moment vector-functions, $E_{1} $ is replaced in FMSWR-case by $H_{1}$, that is, by amplitude of magnetic component of external oscillating EM-field, $J_{E}$ is replaced by the exchange interaction constant 
$J_{H}$, $\gamma_{E}$ by $\gamma_{H}$.
Futher, the linearized equation in \cite{Yearchuck_Yerchak} is considered. It was obtained under the assumption, that the values of oscillating external electric field 
components $E^x, E^y$ in $(\ref{eq1})$ are in experiment greatly less in 
comparison with the value of intracrystalline electric field component 
$E^z$, under analogous assumption relatively the components of total 
electrical spin moment and under additional assumption, that equilibrium distribution of $\vec {S}_0 (z)$ along the 
chain is homogeneous. All the assumptions are entirely correct for IR measurements. The linearized equation was solved  and 
 the relationships for a shape and 
amplitudes of resonance modes and dispersion law were obtained. They are:
\begin{equation}
\label{eq2}
a_{n} = \left\{ {{\begin{array}{*{40}c}
 {-\frac{i \gamma_{E} S \tau^2 E_{1}}{\pi n} \frac{\left[{(\omega_n - \omega
) - \frac{i}{\tau}} \right]}{\left[ {1 + (\omega_{n} - \omega)^2 \tau^2} 
\right]},\,\,n = 1,\,3,\,5,... \hfill} \\
{\,0,\,\,\,\,\,\,\,\,\,\,\,\,\,\,\,\,\,\,\,\,\,\,\,\,\,\,\,\,\,\,\,\,\,\,\,\,\,\,\,\,\,\,\,\,\,\,\,\,\,\,\,\,\,\,\,\,\,\,\,\,\,\,\,\,\,\,\,n = 2,\,4,\,6,...} 
\hfill, \\
\end{array} }} \right. 
\end{equation}
\begin{equation}
\label{eq3}
\nu _n =\nu _0 - A n^2,
\end{equation}
where $n\in N$ including zero, $\nu_{n}$ is a frequency of \textit{n-th} mode, $A$ is 
a material parameter ($A = \frac{2 \pi {a}^{2} S \left|J\right|}{{\hbar L}^{2}} {n}^{2} > 0$).
Here $Re\,a_n $ is proportional to absorption signal, $Im\,a_n 
$ is proportional to dispersion signal.
It was found the followig. 1) Dispersion law (\ref{eq3}) is held true both by IR- and RS-detection of FESWR. 2) The excitation of the only uneven modes in accordance with (\ref{eq2}) takes place. 3) Inversely 
proportional dependence at resonance of the amplitudes of modes on mode number $n$ in accordance with (\ref{eq2}) is also held true, however the only by
IR-detection of FESWR (that is, for the experimental conditions, corresponding to applicability of linearized equation). 4) Splitting of 
Raman active vibration modes is characterized by value of parameter $\mathfrak{A}$, being 
approximately by factor 2 greater, than parameter $\mathfrak{A}$, which characterizes IR FESWR spectra 
(by the frequencies of zero modes reduced by means of linear approximation 
procedure to the same value). The fourth result agrees well with the results, 
reported in \cite{Yearchuck_Doklady}, where the FESWR phenomenon was theoretically described and 
differences of IR FESWR and RS FESWR spectra were predicted. 
To obtain the ratio $J_{E }/J_{H}$ of exchange 
constants the 
values of splitting parameter $\mathfrak{A}^H$ of FMSWR in A-sample reported earlier in \cite{Ertchak_J_Physics_Condensed_Matter}, \cite{Ertchak_Carbyne_and_Carbynoid_Structures} were used. The range of given ratio is 12170 - 15696. To explain in a simple way the 
appearance of two exchange constants, which differ from each other more 
than 4 order for the same chain structures, it was suggested, that EM-field in the matter has the complex charge and consists of elecric and magnetic components, which are real and imagine parts correspondingly. It is well known at the same time the point of view, that EM-field is vector real field, that is, it cannot be characterized by any charges at all.

So, the aim of presented paper is to study the symmetry properties of EM-field characteristics relatively improper rotations in more details and the field theory study of the possibility for the existence of complex charge. 

\section{Algebraic properties of EM-field }
Let us consider the general algebraic properties of EM-field to clarify the symmetry properties first of all of the quantities $\vec {P}$, $\vec {E}$, entering optical Bloch equations and, consequently, to better understand their physical meaning. It is well known, that 
EM-field can be characterized by both contravariant tensor 
$F^{\mu \nu }$ (or covariant $F_{\mu \nu })$ and contravariant pseudotensor 
$\tilde {F}^{\mu \nu }$, which is dual to $F_{\mu \nu }$ (or 
covariant $\tilde {F}_{\mu \nu }$, which is dual to $F^{\mu \nu })$. For example, $\tilde {F}^{\mu \nu }$ is determined by the relation $\tilde {F}^{\mu \nu }=\frac{1}{2}e^{\mu \nu \alpha \beta }F_{\alpha \beta }$, where $e^{\mu \nu \alpha \beta}$ is Levi-Chivita 4-tensor. 
The usage of field tensors and pseudotensors seems to be equally possible 
by description of EM-field and EM-field interaction with the matter. 
So, EM-field pseudotensor can be used to obtain, for instance, the 
field invariants \cite{Landau_Lifshitz_Field_Theory}, \cite {Stephani}. 
The suggestion on equality 
in rights of EM-field tensor and EM-field pseudotensor by 
description of electromagnetic phenomena follows from general consideration 
of the \textit{geometry} of Minkowski space. Really the geometry of any pseudo-Euclidean abstract space of index 1, to which Minkowski space is 
isomorphic \cite{Rashevskii}, determines unambiguously 3 kinds, possessing equal rights, of linear centereuclidian
geometrical objects, that is tensors, pseudotensors and spinors (spin-tensors). Further, in the practice of treatment of experimental results is usally accepted, that electric field strength, electric dipole 
moment and electric polarization vectors are polar vectors. At the same time magnetic 
field strength, magnetic dipole moment and magnetization vectors are considered always to be 
axial vectors. However for the effects, describing optical spectroscopic transitions 
the picture seems to be reverse. It follows directly from the structure of algebraic linear (vector) space, which produce EM-field 
tensors $\left\{{F}^{\mu \nu }(x)\right\}$ and pseudotensors $\left\{\tilde {F}^{\mu \nu }(x)\right\}$, if they are considered to be tensor functions of 4-radius-vector $x$. We keep further formally for the vectors $\vec {E}$ and $\vec {H}$ usual physical meaning of electric field strength and magnetic field strength. At the same time we suggest, that their symmetry 
can be both polar and axial, in particular, $\vec {H}$ can be polar and vector $\vec {E}$ can be axial. The ground for  given idea can be obtained from algebra of linear centereuclidian geometrical objects, that is, in fact from geometry of Minkowsi space. 

Let us define the linear space $\left\langle F,+,\cdot \right\rangle$ over the field of genuine scalars P, the vectors in which are sets of contravariant and covariant EM-field tensors and pseudotensors $\left\{{F}^{\mu\nu}\right\}$, $\left\{\tilde{F}^{\mu\nu}\right\}$, $\left\{{F}_{\mu\nu}\right\}$, $\left\{\tilde{F}_{\mu\nu}\right\}$. It is evident that all the axioms of linear space hold true, that is, if ${F_1}^{\mu \nu}$ and ${F_2}^{\mu \nu}$ $\in F$, 
then 
\begin{equation}\ {F_1}^{\mu \nu} + {F_2}^{\mu \nu} = {F_3}^{\mu \nu} \in F, \end{equation}
and, if 
$F^{\mu \nu }\in F$,
then
\begin{equation}\ \alpha\ F^{\mu \nu } \in F \end{equation} for $\forall \alpha\ \in P$. 
Let us define now the linear algebra $\mathfrak F$ by means of assignment in vector space $\left\langle F,+,\cdot \right\rangle$ of transfer operation $(\ast)$ to dual tensor, using the convolution with Levi-Chivita 
4-tensor $e_{\alpha \beta \mu \nu }$. It is also evident, that in algebra $\left\langle \mathfrak F,+,\cdot, \ast \right\rangle$ the axioms of linear algebra hold true, that is
if ${F_1}^{\alpha \beta}$ and ${F_2}^{\alpha \beta}$ $\in \mathfrak F$, 
then 
\begin{equation} e_{\mu \nu\alpha \beta} ({F_1}^{\alpha \beta} + {F_2}^{\alpha \beta}) = \tilde {(F_1)}_{\mu \nu } + \tilde {(F_2)}_{\mu \nu } \in \mathfrak F, \end{equation}
\begin{equation} ({F_1}^{\alpha \beta} + {F_2}^{\alpha \beta})e_{\alpha \beta \mu \nu }= {F_1}^{\alpha \beta} e_{\alpha \beta \mu \nu }+ {F_2}^{\alpha \beta} e_{\alpha \beta \mu \nu } \in \mathfrak F,\end{equation}
\begin{equation} (e_{\alpha \beta \mu \nu } \lambda {F}^{\mu \nu}) = \lambda (e_{\alpha \beta \mu \nu } {F}^{\mu \nu}) = (e_{\alpha \beta \mu \nu } {F}^{\mu \nu}) \lambda \end{equation} for $\forall \lambda\ \in P$. We can also determine on the space $\left\langle F,+,\cdot \right\rangle$ the functional $\Phi$ by the following relationship \begin{equation} \Phi({F}^{\mu \nu}) \equiv \left\langle {F}^{\mu \nu} | \Phi\right\rangle = {F}^{\mu \nu} \tilde {F}_{\mu \nu }, \end{equation} and 
\begin{equation} \Phi (\tilde {F}^{\mu \nu}) \equiv \left\langle \tilde {F}^{\mu \nu} | \Phi \right\rangle = \tilde {F}^{\mu \nu} {F}_{\mu \nu }. 
\end{equation} 
It is clear, that $\Phi$ on the space $F$ is linear functional. Consequently, the sets ${\Phi({F}^{\mu \nu})}, {\Phi(\tilde {F}^{\mu \nu})}$ of linear functionals on the space $\left\langle F,+,\cdot \right\rangle$ represent themself linear space $\left\langle \Phi,+,\cdot \right\rangle$ over the field of \textit{scalars} $P$, which is dual to space $\left\langle F,+,\cdot \right\rangle$. Therefore, we have 
\begin{equation} \left\langle \Phi,+,\cdot \right\rangle = \left\langle {F}^{\times},+,\cdot \right\rangle. \end{equation} 
It is substantional, that ${F}^{\times}$ is not self-dual. Actually, for example, the vector, which can be built on basis vectors of space $\left\langle F,+,\cdot \right\rangle$ over the field of scalars $P$ with the projections, which are the vectors of space $\left\langle \Phi,+,\cdot \right\rangle$ over the field of scalars $P$, that is, they are pseudoscalars according to definition of functional $\Phi$ and consequently cannot belong to field $P$. So, the starting vector (which is built on basis vectors of space $\left\langle F,+,\cdot \right\rangle$) will be the vector over the field of pseudoscalars and cannot belong to space $F$. Thus, really the space ${F}^{\times}$ is not self-dual. 
It means, that for full physical description of dynamical systems interacting with EM-fied and for description of any physical phenomena at all, where EM-interaction presents, it 
is necessary over the absence of self-duality of the space ${F}^{\times}$ to take always into consideration both the spaces, that is $\left\langle F,+,\cdot \right\rangle$ and $\left\langle \Phi,+,\cdot \right\rangle$. More strictly, known Gelfand triple, including together with spaces $F$ and $\Phi$ Hilbert space has to be taken into account, naturally, if corresponding topology is determined. In the space $\left\langle F,+,\cdot \right\rangle$ over the field of scalars $P$ we can choose 2 physically different subspaces 1)$\left\{{F}^{\mu \nu }\right\}$ and 2)$\left\{\tilde {F}^{\mu \nu }\right\}$ over the scalar field $P$. Analogously can be built the space $\left\langle \tilde F,+,\cdot \right\rangle$ over the field of pseudoscalars $\tilde P$, in which also two new subspaces 3)$\left\{{F}^{\mu \nu }\right\}$ and 4)$\left\{\tilde {F}^{\mu \nu }\right\}$ over the pseudoscalar field $\tilde {P}$ can be choosed. The second case differs from the first case by the following. Symmetry properties of $\vec {E}$ and $\vec {H}$ remain the same, that is, $\vec {E}$ is polar vector, since it is dual vector to antisymmetric 3D pseudotensor, and $\vec {H}$, respectively, is axial. At the same time, the components of vector $\vec {E}$ correspond now to pure space components of electromagnetic field pseudotensor ${\tilde {F}^{\mu \nu}}$, the components of vector $\vec {H}$ correspond to time-space mixed components. Arbitrary element of the third subspace \begin{equation} \alpha {F}^{\mu \nu }(x_1) + \beta {F}^{\mu \nu }(x_2), \end{equation} where $\alpha, \beta \in \tilde {P}$ and $x_1, x_2$ are the points of Minkowski space, 
 represents itself the 4-pseudotensor. Its spatial components, which are the components of antisymmetric 3-pseudotensor, determine dual polar vector $\vec {H}$, while mixed components are the components of 3-pseudovector $\vec {E}$. Therefore, the symmetry properties of the components of vectors $\vec {E}$ and $\vec {H}$ regarding the improper rotations will be inverse to the case 1. It is evident, that in the 4-th case the symmetry properties of the components of $\vec {E}$ and $\vec {H}$ regarding the improper rotations will be inverse to the case 2. 

Given consideration has clear mathematical and physical meaning. It means mathematically, in particular, that if in fixed point of 3D-space the vectors $\vec {E_0}$ and $\vec {H_0}$ have usual transformation properties, that is, they are polar and axial vectors correspondingly, the fields (in vector analysis meaning) of these vectors, that is, the vector-functions, corresponding to given vectors, can have other symmetry properties. For instance, vector-functions $\vec {E}(x) = \vec {E_0} sin x$ and $\vec {H}(x) = \vec {H_0} sin x$ will have opposite symmetry properties in comparison with $\vec {E_0}$ and $\vec {H_0}$ regarding inversion of $x$-coordinate. Physically it means, that interactions of EM-field with the centers, which are 1D-, 2D-, 3D-extended in 3D-space can be quite different between themselwes and especially different in comparison with interaction of EM-field with point centers like to nuclei or electrons.

The main results of presented consideration are the following. 

The sets ${\Phi({F}^{\mu \nu})}, {\Phi(\tilde {F}^{\mu \nu})}$ of linear functionals on the space $\left\langle F,+,\cdot \right\rangle$ represent themself linear space $\left\langle \Phi,+,\cdot \right\rangle$ over the field of \textit{scalars} $P$, which is dual to space $\left\langle F,+,\cdot \right\rangle$, but it is substantial, that given linear space is not self-dual.
Further, given consideration allows to suggest, that free EM-field is 4-fold degenerated regarding improper rotations. The interaction with device (or, generally, with some substance) can relieve degeneracy and can lead, by interaction with extended centers to unusual symmetry of field vector-functions. It is also understandable that spectroscopic transition are not instantaneous, that is the formation of resonance state: field + matter or in particular field + device, or field + matter in general case, which can have very long life time in comparison field mode period. It means in its turn that correct description of spectroscopic transitions can be achieved the only in the frames of fully quantum model, in which field and matter produce two subsystems, possessing equal rights. Realization of concrete field state (one of 4 possible) will, evidently, be determined by symmetry characteristics of EM-field vector-functions in interacting substance. We suggest also, that by interaction of EM-field with the matter the elementar charge carrier size, that is electron size, can be taking into consideration for determination of concept of lengthy centers. Electron size does not exceed the value $~10^{-16}$ cm. Given evaluation follows from the conclusion on applicability of quantum electrodynamics theory up to distances $~10^{-16}$ cm \cite {Ph.Enc.}. Therefore, it seems to be reasonable to suggest, that relatively the electron size any individual atoms can be considered to be 3D extended centers. It means, that even in atomic spectroscopy for dipole moments and electric field strengths the corresponding vector-functions have to be used instead usual vectors and their symmetry relatively improper rotations has to be taken into account. 

It is also understandable, that, if electric field components are 
components of pseudotensor of EM-field (and consequently electric dipole moments are also pseudovectors), the equation of dynamics of optical transitions will have the structure, which is
mathematically equivalent to the structure of the equation for 
dynamics of magnetic resonance transitions (in which magnetic field components are parts of
genuine tensor $F^{\mu \nu }$). In other words, mathematical abstractions in optical Bloch equation
 become, in agreement with results \cite{Yearchuck_Doklady}, \cite{Yearchuck_Yerchak}, real physical meaning, that is really $\vec{E}$ is the part of intracrystalline and 
external electric field, which has axial vector symmetry, $\vec{P}$ is electrical moment, which seems to be built like 
to magnetic moment. It seems to be reasonable along with considered symmetry of EM-field relatively improper rotation, by taking also into account the suggested role of elementar electromagnetic charge size to be space scaling factor, to consider in more details the gauge symmetry of EM-field, which is concerned of charges immediately.

\section{Additional gauge invariance of complex relativistic fields}

It is well known, that free EM-field is vector real field, that is, it cannot be characterized by any real valued charges. We will argue, that EM-field in the matter can be considered to be complex field. In given section we will prove the idea, that for any complex field the conserved quantity, corresponding to gauge symmetry, that is charge, can be in general case also complex. Let $u(x)$ = $\left\{ {\,u_{i}(x) \,} \right\}$, $i = \overline{1,n}$, the set of the functions of some complex relativistic field, that is, scalar, vector or spinor field, given in some space of Lorentz group representations. It is well known, that Lagrange equations for any complex relativistic field can be represented in the form of one matrix relativistic differential equation of the first order in partial derivatives, so called generalized relativistic equation, and analogous equation for field with Hermitian conjugated (complex conjugated in the case of scalar fields) functions $u^{+}(x)$ = $\left\{ {\,u_{i}^{+}(x) \,} \right\}$. They are
\begin{equation}
\label{eq4}
\begin{split}
\raisetag{40pt}
(\alpha_{\mu} \partial_{\mu} + \kappa \alpha_{0}) u(x) = 0\\
-\partial_{\mu} u^{+}(x) \alpha_{\mu} + \kappa u^{+}(x) \alpha_{0} = 0,
\end{split}
\end{equation}
where $\alpha_{\mu},\alpha_{0}$ are matrices with constant numbers' elements and with dimension, which coincides with dimension of corresponding space of Lorentz group representation, realized by $\left\{ {\,u_{i}(x) \,} \right\}$. It is evident, that the transformation
\begin{equation}
\label{eq10}
u'(x) = \beta exp(i \alpha) u(x),
\end{equation}
where $\alpha,\beta\in R$, that is, belong to the set of real numbers, and analogous transformation for Hermitian conjugate functions (or complex conjugate functions in the case of scalar fields) 
\begin{equation}
\label{eq11}
u'^{+}(x) = \beta exp(-i \alpha) u^{+}(x)
\end{equation}
keep Lagrange equations $(\ref{eq4})$ to be invariant. 
So, we have gauge transformation of field functions, which is more general in comparison with usually used. The set $(\beta exp(-i \alpha)$ for all possible $\alpha$, $\beta \in R$ produces the group $\Gamma$, which is direct product of known symmetry group, let designate it by $\mathfrak A$ and multiplicative group $\mathfrak R$ of all real numbers (without zero). Therefore, in the case considered the symmetry group of given complex field asquires additional parameter. So, we will have
\begin{equation}
\label{eq12}
\Gamma(\alpha, \beta) = \mathfrak A (\alpha) \otimes \mathfrak R (\beta)
\end{equation}
Let us find the irreducible representations of the group $\mathfrak R(\beta)$. It has to be taken into account, that the group $\mathfrak R(\beta)$ is abelian group and its irreducible representations $T(\mathfrak R)$ are onedimensional. So, the mapping
\begin{equation}
\label{eq13}
T: \mathfrak R \rightarrow T(\mathfrak R)
\end{equation}
is isomorphism, where
$T(1) = 1$.
Therefore, for $\forall (\beta, \gamma)$ of pair of elements of group $\mathfrak R (\beta)$ the following relationship takes place
\begin{equation}
\label{eq14}
T(\beta, \gamma) = T(\beta) T(\gamma).
\end{equation}

Then, it is easy to show,
that 
\begin{equation}
\label{eq16}
T(\beta) = \beta^ {\frac{\partial{T}}{\partial\gamma}{(1)}}.
\end{equation}
The value ${\frac{\partial{T}}{\partial\gamma}{(1)}}$ can be obtained from the condition
\begin{equation}
\label{eq17}
T(-\beta)= -T(\beta).
\end{equation}
Finally, we have
\begin{equation}
\label{eq18a}
T(\beta) = \beta^{2k+1}= exp[{(2k+1) ln\beta}],
\end{equation}
where $k \in N$. It is evident, that irreducible representations of the group $\Gamma(\alpha, \beta)$ can easy be built, since they are
 direct product of irreducible representations of the groups $\mathfrak A(\alpha)$ and $\mathfrak R(\beta)$, that is, they can be represented in the following form 
\begin{equation}
\label{eq18b}
T(\mathfrak A(\alpha)) \otimes T(\mathfrak R(\beta)) =exp(-i m \alpha) exp[{(2k+1) ln\beta}],
\end{equation}
where
$m, k = 0, \pm1, \pm2, ...$ .
It is clear, that some conserved quantity has to correspond to gauge symmetry of the field, which is determined by the group $\mathfrak R(\beta)$. Really, since generalized relativistic equations are invariant relatively transfomations ($\ref{eq10}, \ref{eq11}$) and variation of action integral with starting Lagrangian is equal to zero, then variation of action integral with transformed Lagrangian will also be zero. Consequently, all the conditions of applicability of Noeter theorem are held true. According to Noeter theorem, the conserved quantity, corresponding to $\nu -th$ parameter ($\nu = \overline{1,k}$) by invariance of field under some $\textit{k}$-parametric symmetry group, is
\begin{equation}
\label{eq19a}
Q_\nu(\sigma) = \int\limits_{(\sigma)}\theta_{\mu \nu}d\sigma_\mu = const,
\end{equation}
where
\begin{equation}
\label{eq19b}
\theta_{\mu \nu} = \frac{\partial{L}}{\partial(\partial_{\mu}u_i)} [\partial_{\rho}u_i X_{\rho \nu} - Y_{i \nu}] - L X_{\mu\nu},
\end{equation}
L is field Lagrangian, $\sigma$ is any spacelike hypersurface. The matrices $X_{\rho \nu}, \text{Y}_{i \nu}$ are determined by matrix representations $\left\|(I_{\nu})_{\mu\ \nu}\right\|$ and $\left\|(J_{\nu})_{i k}\right\|$ of infinitesimal operators of symmetry group in coordinate space and in the space of field functions respectively in accordance with the following relationships
\begin{equation}
\label{eq19c}
X_{\rho \nu} = (I_{\nu})_{\mu \alpha} x_\alpha,
Y_{i \nu} = (J_{\nu})_{i k} u_{k}.
\end{equation}
So, using Noeter theorem, we obtain for 4-vector $\theta_\mu$ the following expression
\begin{equation}
\label{eq19}
\theta_{\mu} = -\frac{\partial{L}}{\partial(\partial_{\mu}u_i)} u_{i} -\frac{\partial{L}}{\partial(\partial_{\mu}u_i^*)} u_{i}^*,
\end{equation}
Components of 4-vector $\theta_\mu$ satisfy to continuity equation
\begin{equation}
\label{eq20}
\partial_{\mu}{\theta_{\mu}} =0.
\end{equation}
Conserving quantity, corresponding to (\ref {eq19}), then is
\begin{equation}
\label{eq21}
Q^{'}_{2} = iQ_{2} = -i \int\theta_{4}d^{3}x.
\end{equation}
So $iQ_{2}$ is 
\begin{equation}
\label{eq22}
iQ_{2} = i\int[\frac{\partial{L}}{\partial(\partial_{\mu}u_i)} u_{i} + \frac{\partial{L}}{\partial(\partial_{\mu}u_i^*)} u_{i}^*]d^{3}x.
\end{equation}
\\

It seems to be evident, that the nature of additional conserved quantity is determined by the type of symmetry. It is clear, that to the additional gauge symmetry will correspond the additional charge. It is seen, that this charge is purely imaginary quantity. Known conserved quantity for any complex field, for instance, for Dirac field, which is consequence of the invariance of field Lagrange equations regarding the transformations $u'(x) = exp(i \alpha)$ and $u(x) u'^{+}(x) = exp(-i \alpha) u^{+}(x)$ is well known electric charge $Q^{E}$ \cite{Bogush}. In general case known conserved part of charge $Q_{1}$ \cite{Bogush} is 
\begin{equation}
\label{eq23}
Q_{1} = -\int[\frac{\partial{L}}{\partial(\partial_{\mu}u_i)} u_{i} - \frac{\partial{L}}{\partial(\partial_{\mu}u_i^*)} u_{i}^*]d^{3}x. 
\end{equation} 
It is real quantity.
So, any complex field can be characterized by complex conserving quantity $Q$, which can be called complex charge \begin{equation} 
\label{eq23a} 
Q = Q_{1} + iQ_{2}. 
\end{equation} 

\section{Discussion and conclusions} 

The presence of complex charge means that 4-vector of current $j_\mu$ for any complex field is complex vector. In its turn, it means, that independently on starting origin of the charges and currents in the matter [they can be result of presence of Dirac field or another complex field] all the characteristics of EM-field in the matter have also to be complex-valued. Given conclusion follows from Maxwell equations, which describe EM-field in the matter and which agree with experimental results well. Therefore, in fact the field of application of Maxwell equations is extended. It means also, that dual electrodynamics, developed by Tomilchick and co-authors, see for instance \cite{Tomilchick}, obtains additional ground. Based field equations in dual electrodynamics \cite{Tomilchick} will be isomorphic to Maxwell equations in complex form. From given isomorphism follows, that $iQ_{2}$-component of charge is so called magnetic charge, that is, $iQ_{2} = iQ^{H}$. Therefore, in the case of EM-field in the matter we will have for full charge the expression
\begin{equation}
\label{eq24}
Q = Q^{E} + iQ^{H}.
\end{equation}
Full charge $Q$ for EM-field in the matter can be called EM-charge. It is clear, that in the case of EM-field in the matter the gauge symmetry group $\Gamma(\alpha,\beta)$ can be represented 
\begin{equation}
\label{eq24a}
\Gamma(\alpha,\beta) = U_{1}(\alpha) \otimes G(\beta),
\end{equation}
 where $U_{1}(\alpha)$ is well known abelian unitary group. It is evident, that in the case of free EM-field 
\begin{equation}
\label{eq24b}
Q_{f} = iQ^{H}.
\end{equation}
The same conclusions are qualitatively held true for quantized fields. It means in particular, that free EM-field quantum, that is photon, possesses by charge, which is imagine. It becomes now physically understandable rather effective interaction of given relativistic particles with matter. Really it is difficult to expect, that the particles with zero rest mass and zero charge can effective interact with matter. Nevetheleess given obvious contradiction with experiment does not even discuss in the physical literature (to our knowledge). It becomes also understandable the mechanism of appearance of real part of charge for EM-field in the matter. The velocity $v$ of EM-field propagation in the matter is less in comparison with the velocity $c$ in vacuum. Consequently hyperbolic rotation of coordinate system takes place, which corresponds to appearance of real component of the charge. The conception of complex characteristics of EM-field in the matter agrees well with all practice of electric circuits' calculation. The formal, but very fruitfull mathematical method for electric circuits' calculation, which uses all complex electric characteristics, see for example \cite {Angot}, becomes now natural explanation.
It is understandable, that $iQ^{H}$ cannot be registered in usual conditions in consequence of its imaginary nature. At the same time, if the interaction, like to exchange interaction, is determined by product of two magnetic charges, they can be registered, for instance, by electric spin wave resonance (ESWR). ESWR was detected experimentally in both ferroelectric (F) and antiferroelectric (AF) cases (see Sec.1) and the work on its more detailed theoretical description is in the progress. At present we can report, that ESW RF and AF EWR.-states description in accordance with \cite{Yearchuck_PL} can be based in the limits, corresponding to the extended models, represented in its ground in \cite{Landau} and being to be the model, developed at present (in press). They both can be very powerfull in comparison with any state /- analysing methods (in the meaning of the detection of interaction between the local centers). We can also add, that conclusion, that EM-field in the matter is complex field opens new possibilities for better understanding for a number of phenomena like to corposants with the aim of its practical use for direct receive of electric energy. 

Let us discuss also the results of Sec.2. Since components of 4-vector $A_\mu$ of EM-field potentials in the matter have to transform by Lorentz group representations, at that by general Lorentz group in general case, we will have 4 possibilities for symmetries of EM-field tensor and its components relatively improper rotations. Really, let components of 4-vector $A_\mu$ transform according to the representation $D({L^{(+)}_+})$ of proper Lorentz group ${L^{(+)}_+}$, coresponding to proper orthochronous transformations, then in the case of improper orthochronous transformations ${L^{(+)}_-}$, proper nonorthochronous ${L^{(-)}_+}$, improper nonorthochronous ${L^{(-)}_-}$ transformations of general Lorentz group $L$ \cite{Fedorov} 4-vector $A_\mu$ will transform correspondingly according to direct product of representations 
\begin{equation} 
\label{eq25} 
\begin{split}
\raisetag{40pt}
D({L^{(+)}_+}) \otimes {D(P)}, \\ D({L^{(+)}_+}) \otimes {D(P^{'})}, \\ D({L^{(+)}_+}) \otimes {D(P)} \otimes {D(P^{'})}, 
\end{split}
\end{equation} 
where $D(P)$ is the representation of space inversion group $P$, $D(P^{'})$ is the representation of time inversion group $P^{'}$, which are subgroups of general Lorentz group $L$. Therefore, 4-vector $A_\mu$ of EM-field along with 
known polar t-even 4-vector can also be axial t-even, polar t-uneven and axial t-uneven 4-vector. It corresponds under the conclusion in Sec.2 to 4 types of EM-field tensors, that is, in other words to partition of
linear space $\left\langle F,+,\cdot \right\rangle$ over the field of genuine scalars and pseudoscalars, the vectors in which are sets of contravariant and covariant electromagnetic field tensors and pseudotensors $\left\{{F}^{\mu\nu}\right\}$, $\left\{\tilde{F}^{\mu\nu}\right\}$, $\left\{{F}_{\mu\nu}\right\}$, $\left\{\tilde{F}_{\mu\nu}\right\}$, on 4 subspaces.

Let us define the complex-valued function $Q(\vec{r},t)$ = $Q_1(\vec{r},t) + iQ_2(\vec{r},t)$ of 2 variables $\vec{r}$ and $t$, where $\vec{r}$ is variable 3D-hypersurface of Minkowski space, $\vec{r} \in R^3$, $t\in (0,\infty)$ is time. Then the follwing theorem takes place.
\textit{Theorem}

Let $ Q_1(\vec{r},t)$, which is determined by 
\begin{equation}
\label{eq26}
Q_1(\vec{r},t) = -\int\limits_{\vec{r}}\left[\frac{\partial{L(t,\vec{r}')}}{\partial \left(\frac{\partial u_i}{\partial x_4}\right)} u_{i} - \frac{\partial{L(t, \vec{r}')}}{\partial \left(\frac{\partial u_i^*}{\partial x_4}\right)} u_{i}^*\right]d\vec{r}'
\end{equation}
and $Q_2(\vec{r},t)$, which is determined analogously, taking into account relationship (\ref {eq22}),
 be continuous functions of both the variables in the range of its definition. Then the complex function $Q(\vec{r},t)$ will be analytical function in complex "plane" of variables $\vec{r}$ and $t$. 

To prove the theorem, it is sufficient to show, that $ReQ(\vec{r},t)$ and $Im Q(\vec{r},t)$ of the function $Q(\vec{r},t) = Q_1(\vec{r},t) + iQ_2(\vec{r},t)$ are satysfying to Cauchy-Riemann conditions, that is the following relations take place
\begin{equation}
\label{cauchy_riemann a}
\frac{\partial Q_1(\vec{r},t)}{\partial \vec{r}} = \frac{\partial Q_2(\vec{r},t)}{\partial t}, 
\end{equation}
\begin{equation}
\label{cauchy_riemann b}
\frac{\partial Q_1(\vec{r},t)}{\partial t} = - \frac{\partial Q_2(\vec{r},t)}{\partial \vec{r}}.
\end{equation}
Let us solve \eqref{cauchy_riemann a} and \eqref{cauchy_riemann b} relative to $Q_2(\vec{r},t)$ (the quantity $Q_1(\vec{r},t)$ is considered to be given). It is apparent, that

\begin{equation}
\label{eq27}
\begin{split}
\raisetag{40pt}
\frac{\partial Q_1(\vec{r},t)}{\partial \vec{r}} = \left.-\left[\frac{\partial{L(t,\vec{r}')}}{\partial \left(\frac{\partial u_i}{\partial x_4}\right)} u_{i} - \frac{\partial{L(t,\vec{r}')}}{\partial \left(\frac{\partial u_i^*}{\partial x_4}\right)} u_{i}^*\right]\right|_{\vec{r}} = \\ \frac{\partial Q_2(\vec{r},t)}{\partial t}. 
\end{split}
\end{equation}
Consequently $Q_2(\vec{r},t)$ is
\begin{equation}
\label{eq28}
Q_2(\vec{r},t) = -\int\limits_{(t)}\left[\frac{\partial{L(\vec{r},t')}}{\partial \left(\frac{\partial u_i}{\partial x_4}\right)} u_{i} - \frac{\partial{L(\vec{r},t')}}{\partial \left(\frac{\partial u_i^*}{\partial x_4}\right)} u_{i}^*\right]dt' + f(\vec{r}). 
\end{equation}
Then, using the relationship \eqref{cauchy_riemann b}, we obtain
\begin{equation}
\label{eq29}
\begin{split}
\raisetag{40pt}
\frac{\partial Q_2(\vec{r},t)}{\partial \vec{r}} = -\frac{\partial}{\partial \vec{r}}\int\limits_{(t)}\left[\frac{\partial{L(\vec{r},t')}}{\partial \left(\frac{\partial u_i}{\partial x_4}\right)} u_{i} - \frac{\partial{L(\vec{r},t')}}{\partial \left(\frac{\partial u_i^*}{\partial x_4}\right)} u_{i}^*\right]dt' \\ + \frac{d f(\vec{r})}{d \vec{r}} = -\frac{\partial Q_1(\vec{r},t)}{\partial t} \\= \frac{\partial}{\partial t} \int\limits_{\vec{r}}\left[\frac{\partial{L(t',\vec{r}')}}{\partial \left(\frac{\partial u_i}{\partial x_4}\right)} u_{i} - \frac{\partial{L(t',\vec{r}')}}{\partial \left(\frac{\partial u_i^*}{\partial x_4}\right)} u_{i}^*\right]d\vec{r}'.
\end{split}
\end{equation}
Hence follows the equation for determination of $f(\vec{r})$
\begin{equation}
\label{eq30}
\begin{split}
\raisetag{40pt}
\frac{d f(\vec{r})}{d \vec{r}} = \frac{\partial}{\partial \vec{r}}\int\limits_{(t)}\left[\frac{\partial{L(\vec{r},t')}}{\partial \left(\frac{\partial u_i}{\partial x_4}\right)} u_{i} - \frac{\partial{L(\vec{r},t')}}{\partial \left(\frac{\partial u_i^*}{\partial x_4}\right)} u_{i}^*\right]dt' \\+ \frac{\partial}{\partial t} \int\limits_{\vec{r}}\left[\frac{\partial{L(t,\vec{r}')}}{\partial \left(\frac{\partial u_i}{\partial x_4}\right)} u_{i} - \frac{\partial{L(t,\vec{r}')}}{\partial \left(\frac{\partial u_i^*}{\partial x_4}\right)} u_{i}^*\right]d\vec{r}'. 
\end{split}
\end{equation} 
So we have 
\begin{equation}
\label{eq31}
\begin{split}
\raisetag{40pt}
 f(\vec{r}) = \int\limits_{(t)}\left[\frac{\partial{L(\vec{r},t)}}{\partial \left(\frac{\partial u_i}{\partial x_4}\right)} u_{i} - \frac{\partial{L(\vec{r},t)}}{\partial \left(\frac{\partial u_i^*}{\partial x_4}\right)} u_{i}^*\right]dt' \\ + \int\limits_{\vec{r}}\left\{ \frac{\partial}{\partial t} \int\limits_{\vec{r}''}\left[\frac{\partial{L(\vec{r},t')}}{\partial \left(\frac{\partial u_i}{\partial x_4}\right)} u_{i} - \frac{\partial{L(\vec{r},t')}}{\partial \left(\frac{\partial u_i^*}{\partial x_4}\right)} u_{i}^*\right]d\vec{r}' \right\}d\vec{r}''. 
\end{split}
\end{equation} 
Consequently 
\begin{equation}
\label{eq32}
\begin{split}
\raisetag{40pt}
\nonumber Q_2(\vec{r},t) = \int\limits_{\vec{r}}\left\{ \frac{\partial}{\partial t} \int\limits_{\vec{r}''}\left[\frac{\partial{L(\vec{r},t)}}{\partial \left(\frac{\partial u_i}{\partial x_4}\right)} u_{i} - \frac{\partial{L(\vec{r},t)}}{\partial \left(\frac{\partial u_i^*}{\partial x_4}\right)} u_{i}^*\right]d\vec{r}' \right\}d\vec{r}'' \\ = - \int\limits_{\vec{r}} \frac{\partial Q_1(\vec{r}'',t)}{\partial t}d\vec{r}''.
\end{split}
\end{equation} 
Then, in suggestion, that dynamic system studied is autonomous, that is
$L(\vec{r},t) = L(\vec{r})$, we will have
\begin{equation}
\label{eq33}
\begin{split}
\raisetag{40pt}
Q_2(\vec{r},t) = \int\limits_{\vec{r}}\left[\frac{\partial{L(\vec{r}'')}}{\partial \left(\frac{\partial u_i}{\partial x_4}\right)} \frac{\partial u_{i}}{\partial t} - \frac{\partial{L(\vec{r}'')}}{\partial \left(\frac{\partial u_i^*}{\partial x_4}\right)} \frac{\partial u_{i}^*}{\partial t}\right]d\vec{r}''.
\end{split}
\end{equation} 
Further, taking into account, that general solution of general relativistic equation is superposition of monochromatic plane waves, which have the view 
$ u_i(t) \sim e^{-i \frac{\mathcal E}{\hbar} t}$, in the simplest case of one plane wave we obtain 
\begin{equation}
\label{eq34}
\begin{split}
\raisetag{40pt}
Q_2(\vec{r},t) = -i \frac{\mathcal E}{\hbar} \int\limits_{\vec{r}} \left[\frac{\partial {L(\vec{r}')}}{\partial \left(\frac{\partial u_i}{\partial x_4}\right)} u_{i}(\vec{r}',t)\right. \\ 
+ \left.\frac{\partial {L(\vec{r}'')}}{\partial \left(\frac{\partial u_i^*}{\partial x_4}\right)} u_{i}^*(\vec{r}',t) \right]d\vec{r}'. 
\end{split}
\end{equation} 
When making a transformation of variable $t \rightarrow ict = x_4$ we finally have 
\begin{equation}
\label{eq35}
\begin{split}
\raisetag{40pt}
Q_2(\vec{r},x_4) = \frac{\mathcal E}{\hbar c} \int\limits_{\vec{r}}\left[\frac{\partial {L(\vec{r}')}}{\partial \left(\frac{\partial u_i}{\partial x_4}\right)} u_{i}(\vec{r}',x_4)\right. \\
+ \left.\frac{\partial {L(\vec{r}')}}{\partial \left(\frac{\partial u_i^*}{\partial x_4}\right)} u_{i}^*(\vec{r}',x_4)\right] d\vec{r}'. 
\end{split}
\end{equation} 
We see, that relationships (\ref{eq22}) and (\ref{eq35}) are coinciding to scaling factor. They will coincide fully, if to make a transformation of parameter $\beta \rightarrow \beta ^{'} = \beta \frac{\mathcal E}{\hbar c}$. The theorem is proved. Given theorem can be considered to be example of independent obtaining of some physical quantities in complex fields.

Therefore, it has been established the partition of
linear space $\left\langle F,+,\cdot \right\rangle$ over the field of genuine scalars and pseudoscalars, the vectors in which are sets of contravariant and covariant EM-field tensors and pseudotensors $\left\{{F}^{\mu\nu}\right\}$, $\left\{\tilde{F}^{\mu\nu}\right\}$, $\left\{{F}_{\mu\nu}\right\}$, $\left\{\tilde{F}_{\mu\nu}\right\}$, on 4 subspaces. It corresponds to appearance of 4 kinds of EM-field potential 4-vectors $A_{\mu}$, which transform according to representations of general Lorentz group with various symmetry relatively improper rotations.

It was found, that the sets ${\Phi({F}^{\mu \nu})}, {\Phi(\tilde {F}^{\mu \nu})}$ of linear functionals on the space $\left\langle F,+,\cdot \right\rangle$ represent themself linear space $\left\langle \Phi,+,\cdot \right\rangle$ over the field of \textit{scalars} $P$, which is dual to space $\left\langle F,+,\cdot \right\rangle$. it is substantial, that given linear space is not self-dual. 

Additional gauge invariance of complex relativistic fields was studied. It has been found, that conserving quantity, corresponding to invariance of generalized relativistic equations relatively the operations of multiplicative group $\mathfrak R$ of all real numbers (without zero) is imagine charge. So it was shown, that complex fields are characterized by complex charges. The irreducible representations of the group $\mathfrak R$ have been found.

It is argued, that EM-field in the matter is complex field and 
that two-parametric group $\Gamma(\alpha,\beta) = U_{1}(\alpha) \otimes \mathfrak R(\beta)$, where $\mathfrak R(\beta)$ is abelian multiplicative group of real numbers (excluding zero), determines the gauge symmetry of EM-field. It is also argued, that free EM-field is characterized by pure imagine charge.
We believe, that all the conclusions relatively additional gauge invariance are held also true for neutrino field in the case of weak couplings.
The theorem on analicity of complex-valued function $Q(\vec{r},t)$ = $Q_1(\vec{r},t) + iQ_2(\vec{r},t)$ of 2 variables 
$\vec{r}$ and $t$ is proved.

The results obtained agree well with experimental and previous theoretical data, reviewed shortly in Sec.1. 

\section*{Aknowledgement}
Authors are grateful to Y.Yerchak for discussions and the help in the work.

\end{document}